\documentclass[12pt]{iopart}

\usepackage{iopams}

\usepackage{graphicx}

\usepackage{graphicx}
\usepackage{epstopdf, epsfig}

\usepackage{wrapfig}
\usepackage{lipsum}

\usepackage[utf8]{inputenc}
 
\usepackage{dblfloatfix}

\usepackage{dcolumn}
\usepackage{bm}
\usepackage{amsfonts}

\usepackage{color}

\usepackage{array}

\usepackage{setspace}

\usepackage{graphicx}
\usepackage{ragged2e}

\usepackage[font=small,labelfont=bf]{caption}

\usepackage{enumitem}

\usepackage{multirow}

\usepackage[superscript,biblabel]{cite}
\usepackage{textcomp}

\usepackage{tabularx}
\usepackage{adjustbox}
\usepackage{booktabs}

\usepackage{multirow}
\usepackage[table,xcdraw]{xcolor}

\usepackage[utf8]{inputenc}

\usepackage[font=scriptsize,labelfont=bf]{caption}

\usepackage{enumitem}

\definecolor{bl}{rgb}{0.5, 0, 0}
\definecolor{cobalt}{rgb}{0.0, 0.28, 0.67}
\definecolor{darkcerulean}{rgb}{0.03, 0.27, 0.49}
\definecolor{egyptianblue}{rgb}{0.06, 0.2, 0.65}

\usepackage{hyperref}
\usepackage{xcolor}
\hypersetup{
    colorlinks = true,
    linkcolor = blue,
    citecolor = blue,
    urlcolor = black
}

\usepackage[superscript,biblabel]{cite}

\usepackage{graphicx}

\usepackage[T1]{fontenc}
\usepackage[utf8]{inputenc}

\begin{document}

\title[Nasal spray usage]{Numerical evaluation of spray position for improved nasal drug delivery}

\author[Basu et al.]{\footnotesize{Saikat Basu$^{1, *}$, Landon T Holbrook$^2$, Kathryn Kudlaty$^3$, Olulade Fasanmade$^3$, Jihong Wu$^2$, Alyssa Burke$^2$, Benjamin Langworthy$^4$, Zainab Farzal$^3$, Mohammed Mamdani$^3$,  William D Bennett$^2$, Jason P Fine$^4$,  Brent A Senior$^3$, Adam M Zanation$^3$, Charles S Ebert, Jr.$^3$, Adam J Kimple$^3$, Brian D Thorp$^3$, Dennis O Frank-Ito$^5$, Guilherme JM Garcia$^6$, and Julia S Kimbell$^3$}}
\vspace{0.5cm}

\address{$^1$~Department of Mechanical Engineering, South Dakota State University, Brookings, SD 57007, United States\\
         $^2$~Center for Environmental Medicine, Asthma and Lung Biology, University of North Carolina, Chapel Hill, NC 27599, United States\\
         $^3$~Department of Otolaryngology / Head and Neck Surgery, School of Medicine -- University of North Carolina, Chapel Hill, NC 27599, United States\\
	     $^4$~Department of Biostatistics, University of North Carolina, Chapel Hill, NC 27599, United States\\	
	     $^5$~Department of Head and Neck Surgery \& Communication Sciences, Duke University Medical Center, Durham, NC 27708, United States\\
	     $^6$~Joint Department of Biomedical Engineering, Medical College of Wisconsin and Marquette University, Milwaukee, WI 53226, United States
	     }
\ead{$^*$\,Saikat.Basu@sdstate.edu}


\date{}





\begin{abstract}
\textbf{Topical intra-nasal sprays are amongst the most commonly prescribed therapeutic options for sinonasal diseases in humans. However, inconsistency and ambiguity in instructions show a lack of definitive knowledge on best spray use techniques. In this study, we have identified a new usage strategy for nasal sprays available over-the-counter, that registers an average 8-fold improvement in topical delivery of drugs at diseased sites, when compared to prevalent spray techniques. The protocol involves re-orienting the spray axis to harness inertial motion of particulates and has been developed using computational fluid dynamics simulations of respiratory airflow and droplet transport in medical imaging-based digital models. Simulated dose in representative models is validated through \textit{in vitro} spray measurements in 3D-printed anatomic replicas using the gamma scintigraphy technique. This work breaks new ground in proposing an alternative user-friendly strategy that can significantly enhance topical delivery inside human nose. While these findings can eventually translate into personalized spray usage instructions and hence merit a change in nasal standard-of-care, this study also demonstrates how relatively simple engineering analysis tools can \textit{revolutionize} everyday healthcare.}\\
\vspace{-0.4cm}
\flushright\noindent(175 words)\\
\end{abstract}
\vspace{-0.9cm}

\newpage

\section{Introduction}

\begin{wrapfigure}{r}{0.6\textwidth} 
\vspace{-1.35cm}
\begin{center}
\includegraphics[width=0.6\textwidth]{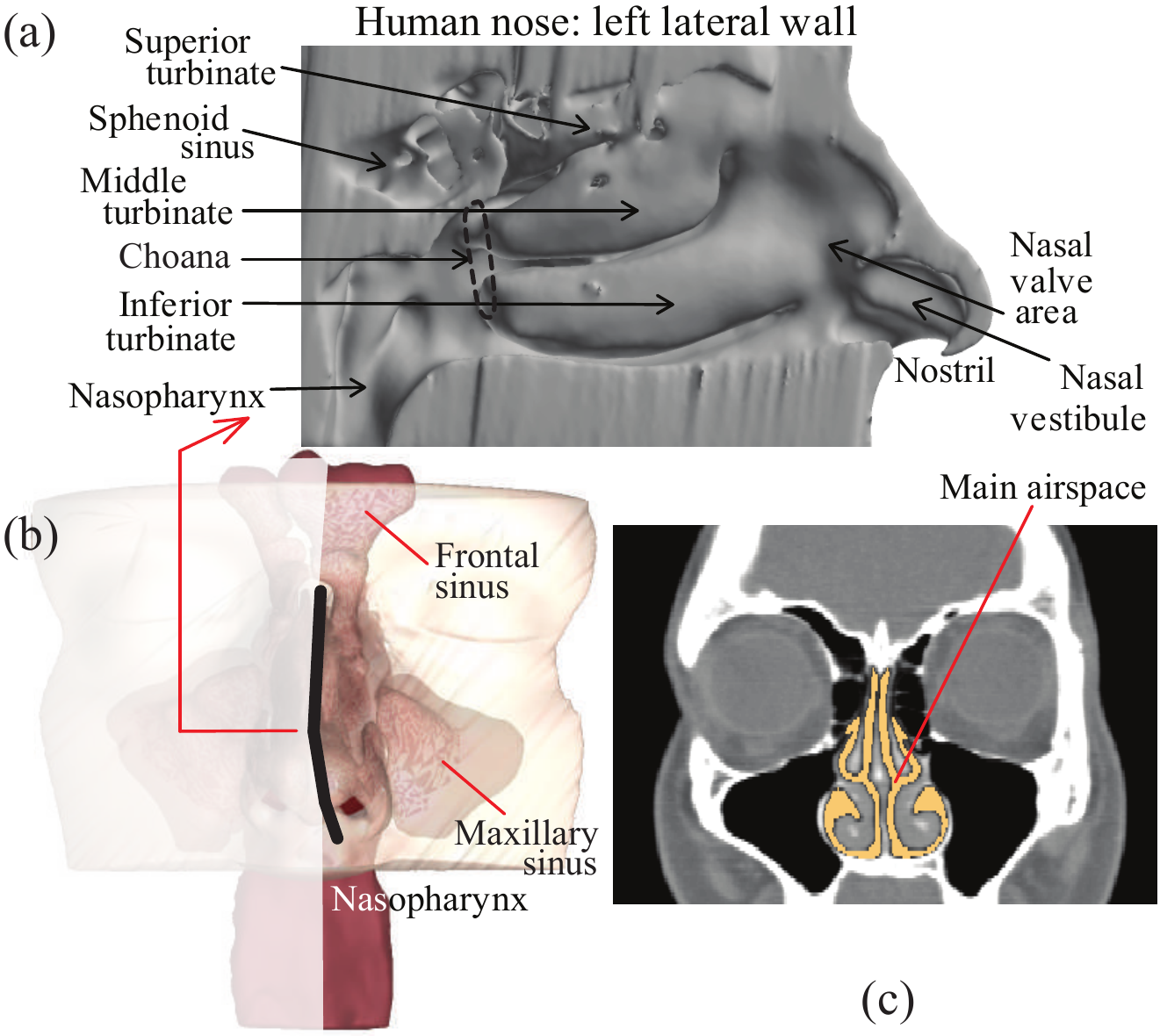}
\vspace{-0.5cm}
\caption{\scriptsize{(a) Anatomic features inside a human nose, as viewed on an invasive cut-away. (b) Position of the cut-away section, marked by dark line. (c) Representative coronal section, with the main nasal passage shown in lighter color.}}
\label{fig:human-nose}
\end{center}
\vspace{-0.75cm}
\end{wrapfigure}

Inside of our nose is structurally and physiologically complex (e.g.~see Figure~\ref{fig:human-nose}). It comprises the main intra-nasal passage, the mucous membrane, the ciliary hair-like cells, the mucosal drainage fluid circulating along the internal walls, and the adjoining sinus cavities of various shapes and sizes\cite{Doorly:2008gh, proctor2017comparative}. Occlusion of the sinus chambers with mucus is associated with many nasal ailments, such as chronic rhinosinusitis \cite{benninger2003ohns}. While surgical treatments essentially focus on enlarging the opening to the sinus chambers, such procedures can be cost-prohibitive, have associated risks, and are mostly reserved for medically refractory diseases. As a first line of treatment, physicians often recommend topical sprays \cite{parikh2001rhino, rosenfeld2007}, with the rationale that these topical drugs will reduce inflammation at the diseased sites and assist in resolving the occlusion and re-establishing natural drainage. However, while such sprays do rank amongst the most commonly used therapeutics, the efficacy of the drugs can be highly patient-specific and there is no well-defined protocol to ensure that specific dosage would reach the intended intra-nasal target sites.

Transport of topical drugs inside the nose encounters a number of challenges, namely the airway tortuosity, the sweeping effect of mucociliary drainage, and a lack of consistent usage protocol for the medical devices employed in drug application (primarily owing to the inter-subject heterogeneity in internal anatomic geometry). While optimizing the trajectories for topical nasal drug transport by experimental trials in real human subjects is improbable; with advancement of computational tools, there has been a significant push to obtain numerically simulated predictions of respiratory flow physics and transport therein; see e.g.\cite{zhao2011plos, inthavong2009ct, kimbell2019lsm}. Of interest are nasal spray simulation studies on \textit{in silico} models, re-constructed from medical imaging, to measure drug delivery along the nasal passages\cite{inthavong2011jas}, in the sinuses\cite{basu2017num}, and on the effects of surgical alterations of the anatomy on nasal airflow\cite{frank2013ifar, brandon2017jama, tracy2019ifar2} as well as on topical transport of drugs\cite{kimbell2018rdd, farzal2018ifar, paper3, basu2019aps}. The latter addresses the role of airway channel's shape in the context of airflow-droplet interactions. Notably, while using medical devices like sprayers, which are inserted at the nostril, the anterior airway geometry gets altered. To simplify the situation though, computational results \cite{basu2017num} suggest that such initial perturbations do not greatly change or adversely affect the eventual drug deposits at the diseased sites. 

Despite the abundance of computational research on nasal drug delivery, there is a distinct lack of articulate instructions for guidance on what could be the ``best'' way to use the commercially available sprayers. First, numerical studies often do not use a realistic distribution of droplet sizes while simulating topical sprays. Focusing on specific droplet diameters is resourceful while studying the detailed nuances of transport characteristics in that size range; however this somewhat limits the applicability of the subsequent findings while predicting the performance of real sprays, which have a wide variability of droplet sizes in each spray shot. Secondly, the inter-subject anatomic variations also render it difficult to identify a generic spray orientation that can work for all and ensures maximal delivery of drugs at the diseased locations inside the nose.

In this study, we have numerically tracked the transport of therapeutic particulates from over-the-counter nasal sprays via inhaled airflow. The computational fluid dynamics (CFD) models of droplet transport and the \textit{in silico} prediction of their deposition sites along the nasal airway walls have been compared with \textit{in vitro} spray experiments in 3D-printed solid replicas of the same anatomic reconstructions. We have proposed a new strategy of nasal spray usage and the recommendation is supported by a significant improvement in target site particulate deposition (TSPD), when compared to the prevalent spray use techniques. The study also expounds\cite{leong2010rhino, zubair2012jmbe, burrowes2017sbm} on the potential of CFD as a tool in nasal ailment treatment and subject-specific prognosis, and can contribute to the emergence of non-invasive personalized therapeutics and treatment strategies. 


\section{Methods}

\subsection{Anatomic reconstructions}
We have used de-identified computed tomography (CT) data from three pre-surgery chronic rhinosinusitis (CRS) patients; under approval from the Institutional Review Board (IRB) at the University of North Carolina at Chapel Hill. Subject 1 was a 41 year-old Caucasian male (body weight 88.0 kg, body mass index 25.3); subject 2 was a 70 year-old Caucasian male (body weight 67.5 kg, body mass index 24.8); and subject 3 was a 24 year-old Caucasian female (body weight 93.1 kg, body mass index 32.6). Medical-grade CT scans of the subjects' nasal airways were used to re-construct digital models through thresholding of the image radiodensity, with a delineation range of -1024 to -300 Hounsfield units for airspace\cite{borojeni2017, basu2017num}, complemented by careful manual editing of the selected pixels for anatomic accuracy. As part of that process, the scanned DICOM (Digital Imaging and Communications in Medicine) files for each subject were imported to the image processing software Mimics\textsuperscript{TM} v18.0 (Materialise, Plymouth, Michigan). For this study, we subsequently considered each side of the nose in the \textit{in silico} models as a distinct nasal passage model, while studying the droplet  transport properties when the spray nozzle was placed on that side:~(a) subject 1's right side constituted nasal passage model 1 (NPM1) and his left side was nasal passage model 2 (NPM2); (b) subject 2's left side was nasal passage model 3 (NPM3); and (c) subject 3's right side was nasal passage model 4 (NPM4) and her left side was nasal passage model 5 (NPM5). Note that subject 2's right-side anatomy did not exhibit a direct access to the diseased intra-nasal targets from outside of the nostril and was not selected for this study. This had to do with the scope of our study design; for details, see Section~\ref{s:StudyDesign}. Also refer to the discussion section for follow-up comments.

To prepare the \textit{in silico} anatomic models for numerical simulation of the inhaled airflow and the sprayed droplet transport therein, the airway domain was meshed and spatially segregated into minute volume elements. The meshing was implemented by importing the Mimics-output in stereolithography (STL) file format to ICEM-CFD\textsuperscript{TM} v18 (ANSYS, Inc., Canonsburg, Pennsylvania). Following established protocol\cite{basu2017jampdd, basu2017num}, each computational grid comprised approximately 4 million unstructured, graded tetrahedral elements; along with three prism layers of approximately 0.1-mm thickness extruded at the airway-tissue boundary with a height ratio of 1.



\begin{table}[b]
\caption*{List of acronyms.}\label{Table0}
\vspace{-0.2cm}
\centering
\footnotesize
\begin{adjustbox}{width=0.6\textwidth}
\begin{tabular}{|
>{\columncolor[HTML]{EFEFEF}}l |l|}
\hline
\cellcolor[HTML]{C0C0C0}Full name & \cellcolor[HTML]{C0C0C0}Acronym \\ \hline
NPM                               & Nasal Passage Model             \\ \hline
TSPD                              & Target Site Particulate Deposition / Delivery \\ \hline
CT                                & Computed Tomography          \\ \hline
CRS                               & Chronic Rhinosinusitis          \\ \hline
OMC                               & Ostiomeatal Complex             \\ \hline
CFD                               & Computational Fluid Dynamics    \\ \hline
DICOM                             & Digital Imaging and Communications in Medicine  \\ \hline
STL                               & Stereolithography    \\ \hline
ROI                               & Region of Interest    \\ \hline
NPD                               & Nozzle Positioning Device    \\ \hline
CU                                & Current Use (\textit{spray usage protocol})    \\ \hline
LoS                               & Line of Sight (\textit{spray usage protocol})    \\ \hline
\end{tabular}
\end{adjustbox}
\end{table}


\subsection{Inspiratory airflow and sprayed droplet transport simulations}

Laminar steady-state models work as a reasonable approximation while modeling comfortable resting to moderate breathing\cite{kelly2000jap, kelly2000, kimbell2019lsm, xi2008ijhmt, shanley2008it}. Furthermore, with our simulations focusing on a single cycle of inspiration, steady state flow conditions were adopted as a feasible estimate. Based on the principle of mass conservation (\emph{continuity}), and assuming that the airflow density stays invariant (\textit{incompressibility}), we have
\begin{equation}\label{e:continuity}
\nabla \cdot \mathbf{u} = 0,
\end{equation}
with $\mathbf{u}$ representing the velocity field for the inspired air. Conservation of momentum under steady state flow conditions leads to the modified Navier-Stokes equations:~
\begin{equation}\label{e:NS}
\rho\left(\mathbf{u} \cdot \nabla \right)\mathbf{u} = -\nabla p + \mu {\nabla}^2 \mathbf{u}+\rho\mathbf{b}.
\end{equation}
Here $\rho = 1.204$ kg/m$^3$ represents the density of air, $\mu = 1.825\times10^{-5}$ kg/m.s is air's dynamic viscosity, $p$ is the pressure in the airway, and $\mathbf{b}$ stands for accelerations induced by different body forces. To simulate the airflow, equations~(\ref{e:continuity}) and (\ref{e:NS}) were numerically solved 
through a finite volume approach, in the inspiratory direction. The computational scheme on ANSYS Fluent\textsuperscript{TM} v14.5 employed a segregated solver, with SIMPLEC pressure-velocity coupling and second-order upwind spatial discretization. Solution convergence was obtained by minimizing the flow residuals (viz.~mass continuity$\,\sim\mathcal{O}(10^{-2})$, velocity components$\,\sim\mathcal{O}(10^{-4})$), and through stabilizing the mass flow rate and the static outlet pressure at the nasopharynx of the digital models. A typical simulation convergence run-time with 5000 iterations clocked approximately 10 hours, for 4-processor based parallel computations executed at 4.0 GHz speed.

\begin{table}[b]
\caption{Parameters for inhalation airflow.}\label{Table0b}
\vspace{-0.35cm}
\begin{center}
\includegraphics[width=0.75\textwidth]{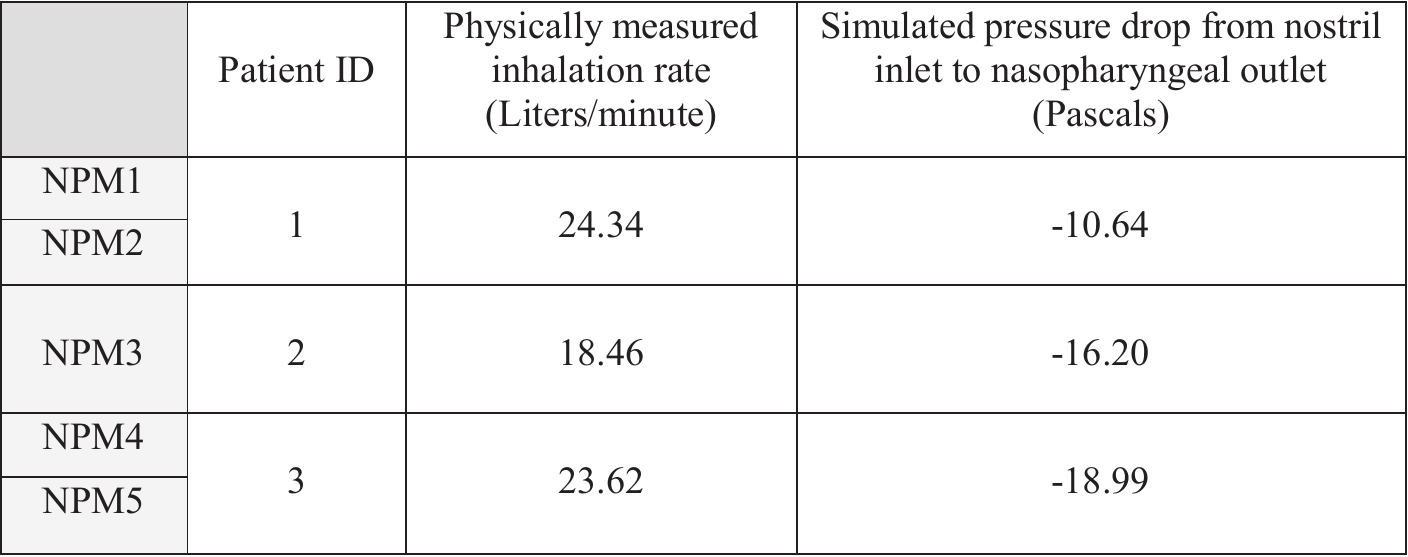}
\end{center}
\vspace{-0.5cm}
\end{table}

The numerical solutions implemented the following set of boundary conditions: (1) zero velocity at the airway-tissue interface i.e.~the tissue surface lining the sinonasal airspace (commonly called \emph{no slip} at the walls), along with ``trap'' boundary conditions for droplets whereby a droplet comes to rest after depositing on the wall; (2) zero pressure at nostril planes, which were the pressure-inlet zones in the simulations, with ``escape'' boundary condition for droplets that allowed outgoing trajectories to leave the airspace through the nostril openings; and (3) a negative pressure at the nasopharyngeal outlet plane, which was a pressure-outlet zone, also with an ``escape'' boundary condition for droplets. The negative nasopharyngeal pressure generated an inhalation airflow rate within $\pm~5-6\%$ of the subject-specific measurement of resting breathing, obtained using LifeShirt\textsuperscript{\textregistered} vests\cite{wilhelm2003bm} that tracked chest compression/expansion during breathing, and accordingly quantified the inhalation rates (see Table~\ref{Table0b}).

After simulating the airflow, sprayed droplet dynamics were tracked through discrete phase particle transport simulations in the ambient airflow, and the corresponding Lagrangian tracking estimated the localized deposition along the airway walls through numerical integration of the following transport equations\cite{fluent14point5}:
\begin{equation}
\frac{d \mathbf{u_d}}{dt} = \frac{18\mu}{d^2 \rho_d }\frac{C_D Re}{24}(\mathbf{u}-\mathbf{u_p})+\mathbf{g}\left(1- \frac{\rho}{\rho_d}\right) + \mathbf{F_B}.
\end{equation}
The parameters here are $\mathbf{u_d}$, representing the droplet velocity; along with $\mathbf{u}$ as the airflow field velocity, $\rho$ and $\rho_d$ respectively as the air and droplet densities, $\mathbf{g}$ as the gravitational acceleration, $\mathbf{F_B}$ as any other additional body forces per unit droplet mass (as for example, Saffman lift force that is exerted by a typical flow-shear field on small particulates transverse to the airflow direction), and $18\mu\, C_D\, Re\,(\mathbf{u}-\mathbf{u_d})/24(d^2 \rho_d )$ quantifies the drag force contribution per unit droplet mass. Here, $C_D$ is the drag coefficient, $d$ is the droplet diameter, and $Re$ represents the relative Reynolds number.

Mean time step for droplet tracking was in the order of $10^{-5}$ sec., with the minimum and maximum limits for the adaptive step-size being $\sim\mathcal{O}(10^{-10})$ sec.~and $\sim\mathcal{O}(10^{-3})$ sec., respectively. Also note that the solution scheme posits the particulate droplets to be large enough to ignore Brownian motion effects on their dynamics. Post-processing of the simulated data laid out the spatial deposition trends, which were then tallied against \textit{in vitro} observations.


\subsection{3D printing and physical experiments}

\begin{figure}[t]
\begin{center}
\includegraphics[width=1.0\textwidth]{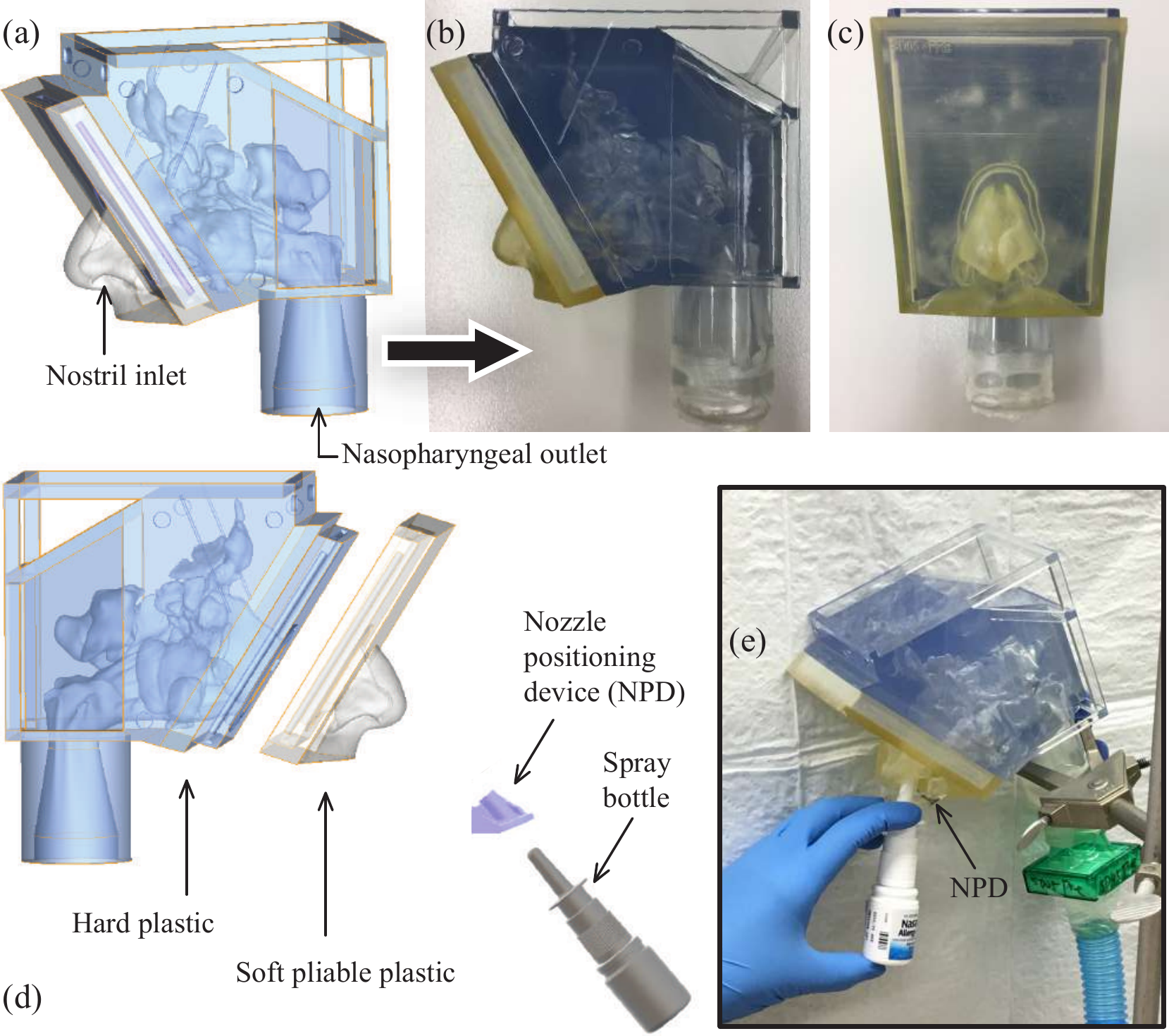}
\caption{(a) \textit{In silico} model: CT-based digital reconstruction of subject 1's airway. Panels (b) and (c) respectively show the sagittal and coronal views of the solid 3D-printed replica of the digital model. Note that the solid models comprise a soft outer nose (to mimic the pliability of a real nose) and a posterior hard plastic part. The anterior and posterior 3D-printed components in each model were designed to fit snugly together. Panels (d) and (e) depict the experimental setup for \textit{in vitro} measurement of sprayed deposits in anatomic solid models.}\label{f:3Dprinting}
\end{center}
\vspace{-0.5cm}
\end{figure}

To assess the reliability of numerically predicted topical deposition vis-à-vis physical experiments, 3D-printed anatomic replicas were generated for subject 1's airway and hence included both NPM1 and NPM2. The posterior parts of the solid models were made from the stereolithography material Watershed\textsuperscript{\tiny\textregistered} (DSM Somos, Elgin, Illinois). Post-digitization, the printing job of the posterior component was sub-contracted to ProtoLabs (Morrisville, North Carolina). Printing of the anterior soft plastic part on a Connex3\textsuperscript{TM} 3D printer was done by Ola Harrysson’s group at North Carolina State University (at the Edward P Fitts Department of Industrial and Systems Engineering), using polymer inkjetting process on Tangogray FLX950 material. See Figure~\ref{f:3Dprinting}(a)-(c) for representative pictures of a digitized model and the corresponding 3D replica.

\subsubsection{Recording deposits through gamma scintigraphy:}
Intra-nasal topical delivery was tracked through \textit{in vitro} examination of mildly radioactive spray deposits in the 3D-printed anatomic replicas. To ensure that the spray axis orientation and nozzle location aligned with the corresponding simulated spray parameters, we used specially designed nozzle positioning devices (NPD) inserted at the nostril. The spray bottle was fitted into the NPD, while administering the spray via hand-actuation.  For each sample test, a bottle of commercial nasal spray Nasacort\textsuperscript{TM}~was labeled with a small amount of radioactive Technetium (Tc99m) in saline. At the time of dispensing the spray shots, a vacuum line controlled by a flow-valve was used to set up inhalation airflow through the model, and the flow rate was commensurate with the subject-specific breathing data (Table~\ref{Table0b}). Corresponding setup is in Fig.~\ref{f:3Dprinting}(d)-(e). Four independent replicate runs of each spray experiment were conducted, followed by compilation of the means and standard deviations of the drug deposits along the inner walls of the solid models. The topical deposition was proportional to the radioactive signals emitted from the spray solution traces that deposited inside a solid model and was quantifiable through image-processing of the scintigraphy visuals, collected using a BodyScan (MieAmerica, Forest Hills, IL) 400-mm width by 610-mm height 2D gamma camera. The pixel domain was 256$\times$256, with an image acquisition time of 3 minutes; and one pixel equated to a  Cartesian distance of 2.38 mm in the digital and 3D models.

\begin{figure}[t]
\vspace{-0.15cm}
\begin{center}
\includegraphics[width=1.0\textwidth]{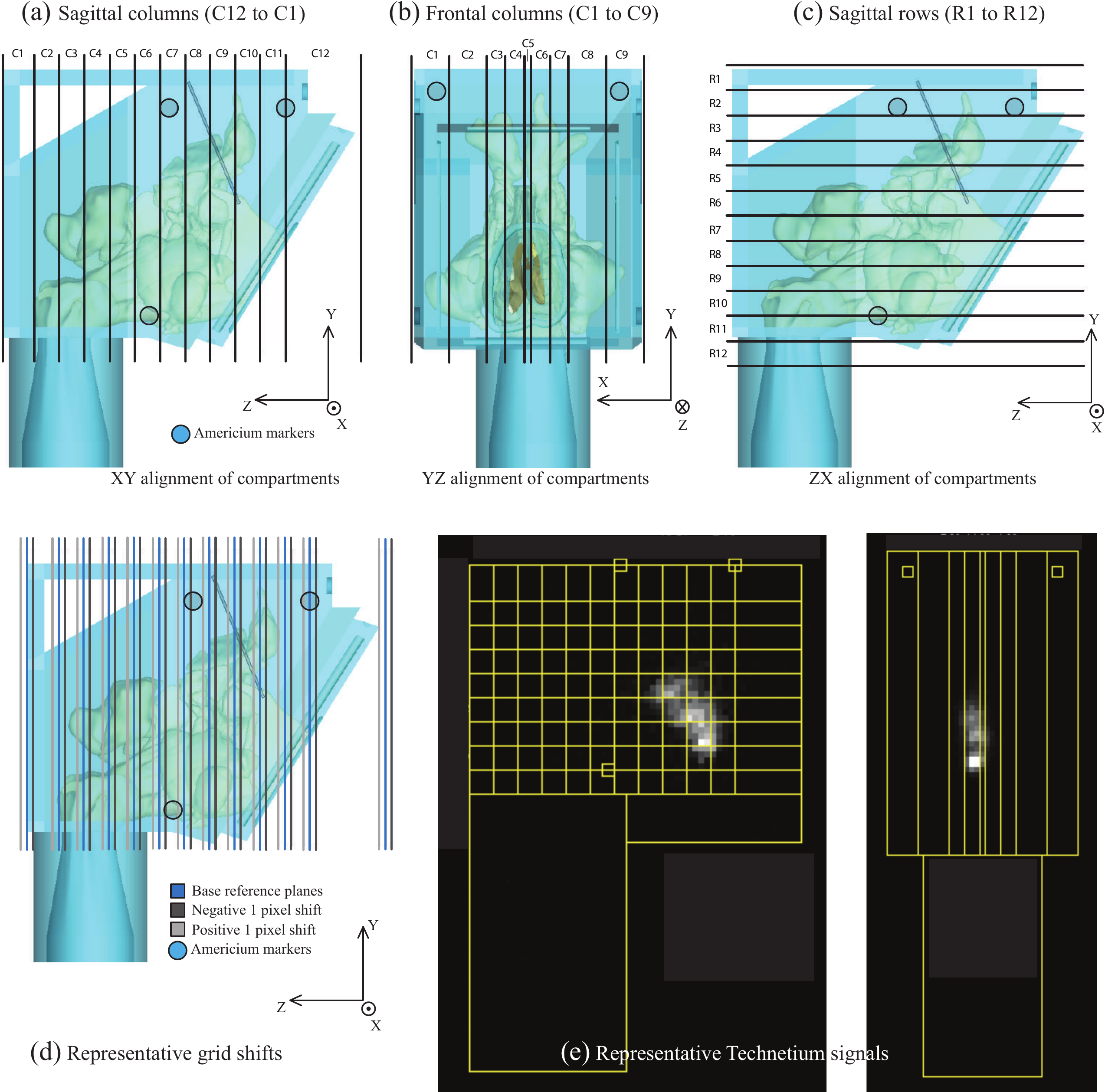}
\caption{Panels (a), (b), and (c) depict the gridline schematic on NPM1 and NPM2, that is used to extract the deposition fractions from the gamma scintigraphy-based quantification of the sprayed deposits in the solid replicas. The models are respectively segregated into 3 sets of compartments: sagittal columns, frontal columns, and sagittal rows. Panel (d) shows the perturbation of the base gridline by 1 pixel. Representative Technetium signals are in panel (e). Note:~In regard to the axis system, the circle with solid dot implies out-of-plane direction from this page, the circle with cross signifies into-the-plane of this page.}\label{f:gridlines}
\end{center}
\vspace{-0.5cm}
\end{figure}

\subsubsection{Model segmentation for comparison with numerical data:}\label{s:gridlines}
To facilitate the comparison between the numerical predictions on droplet deposition and the physical observation of gamma scintigraphy signals in the corresponding solid replica, we segregated NPM1 and NPM2 into virtual segments oriented along three different directions. Figure~\ref{f:gridlines} lays out the Cartesian coordinate directions for the 3D space. X was perpendicular to the sagittal plane traversing from left to right sides of the nasal models (with the model head facing forward), Y was perpendicular to the axial plane traversing from inferior to superior aspects of the models, and Z was perpendicular to the coronal plane traversing from anterior to posterior aspects of the models. The virtual segments were oriented along the XY (coronal), YZ (sagittal), and ZX (axial) planes. Parallel to the XY coronal plane, the models contained 12 segments (named, C$12$ --  C$1$ $\Rightarrow$ sagittal columns); there were 9 compartments (C$1$ -- C$9$ $\Rightarrow$ frontal columns) parallel to the YZ sagittal plane, and there were 12 compartments (R$1$ -- R$12$ $\Rightarrow$ sagittal rows) parallel to the ZX axial plane (see Figure~\ref{f:gridlines}).

For each compartment, the particulate deposition fraction predicted from the simulation was compared with the deposition fraction measured based on gamma signals of the deposited particulates in the corresponding compartment of the 3D-printed model. To achieve this, signals emitted from the solution traces, that settled along the airway walls, were subjected to image processing analysis. Therein, by superimposing the compartmental grid on the radio-images, the signals were extracted from each compartment. In order to align the grid on the image in a manner consistent with the virtual model, three inset discs were designed as reference points on the outer surface of the virtual and 3D-printed models. Americium sources from commercial in-home smoke detectors were inserted into the insets as reference points on the 3D-model and a radio-image was recorded. For the analysis, the scintigraphy images were processed using ImageJ\cite{schneider2012nature} by constructing a region of interest (ROI) referenced to the fixed Americium sources. Care was taken to align the emitted visual signals with similar reference regions within the superimposed grid. This was done via manual visualization to achieve a best fit of signal intensity within reference regions. The grid compartment planes positioned using this visual best-fit technique were designated as ``reference planes''. Given the nature of the radioactive signals and the resolution of the radio-image, some signal intensity resided outside of reference regions even while using best-fit practices. A reasonable fit could be obtained by shifting the image by one pixel in either direction (positive shift / negative shift). In order to account for this variation, alternative plane positions (see Figure~\ref{f:gridlines}(d)) were created by shifting the reference planes one pixel along the positive and negative axes for each set of Cartesian planes. These three sets of compartment planes were positioned in the \textit{in silico} modeling software using the measured distances from the reference regions. The corresponding Cartesian coordinates of these planes were used to assign droplet deposition locations from the computational simulations to grid compartments, for comparison with the \textit{in vitro} model. In these comparisons, we left out the deposits in the anterior nose (from the CFD data as well as the physical recordings) in order to negate the bright radiation signal coming from that zone in the experimental deposits; and focused only on measurements from the posterior parts of the respective models. Note that the anterior nose in an \textit{in silico} model is in fact the removable soft pliable anterior part in the corresponding 3D print (e.g.~see Figure~\ref{f:3Dprinting}).

\begin{figure}[t]
\vspace{-0.15cm}
\begin{center}
\includegraphics[width=1.0\textwidth]{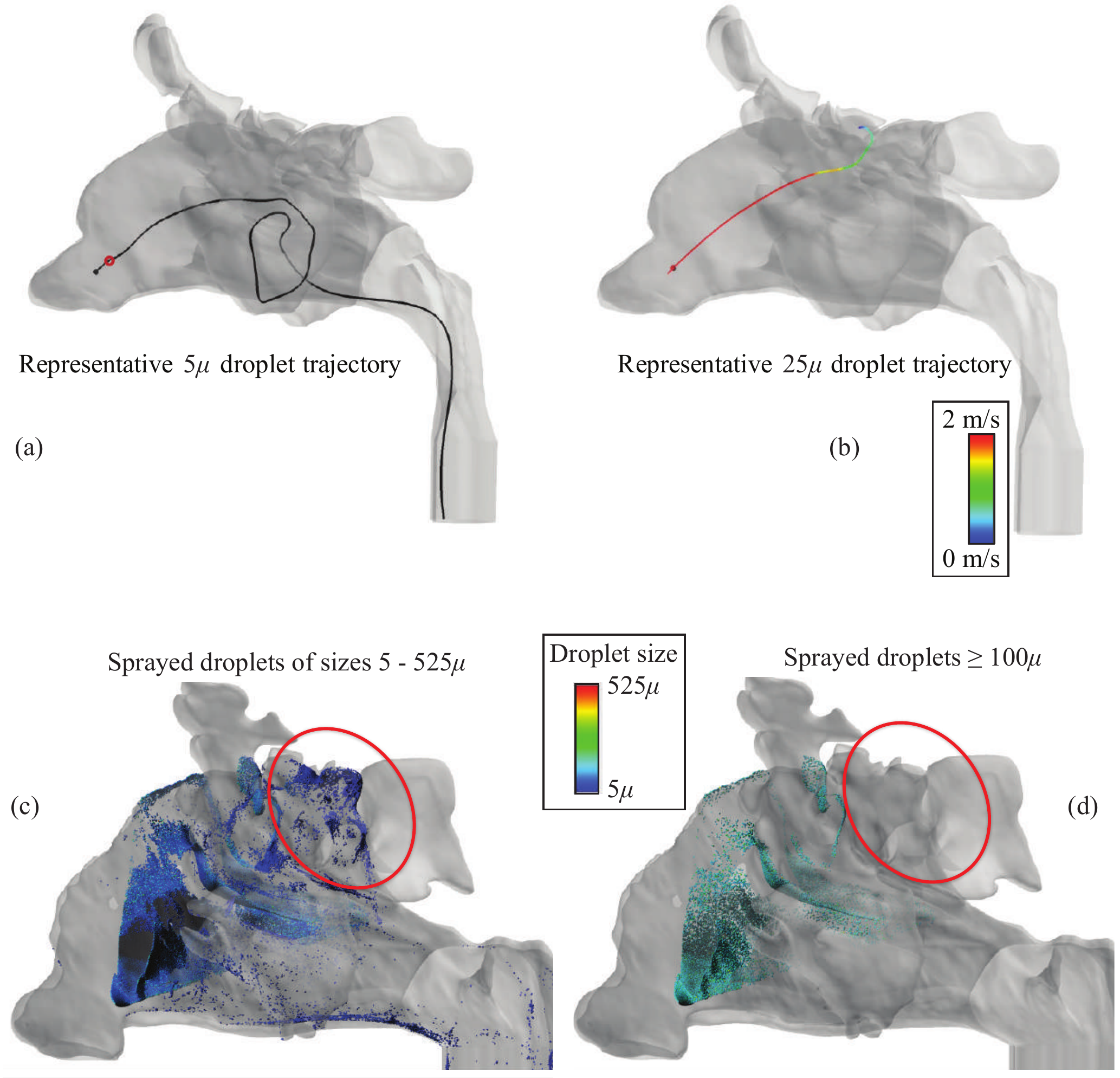}
\caption{Comparison of representative trajectories for a 5$\mu$ droplet and a 25$\mu$ droplet in a sample sinonasal airspace. In panel (a), the smaller droplet has weaker inertial momentum and the ambient airflow streamline takes over its motion much earlier than that in case of a heavier droplet like the one in panel (b), where the inertial momentum of the 25$\mu$ droplet persists longer. The small red circle in (a) depicts the point where the inertial momentum gets overwhelmed by the fluid streamline. Evidently, owing to smaller inertia, the droplets with smaller diameters get predominated by the airflow streamlines earlier than the bigger droplets. This results in a better penetration and spread of sprayed droplets in the nasal airspace, as shown in panel (c), for a different nasal model. On the contrary, spray shots with exclusive share of bigger droplets (e.g.~$\ge$ 100$\mu$ here) tend to follow their initial inertial trajectories, without much effect of the airflow streamlines on their paths, and deposit along the anterior walls of the nasal airspace, as depicted in panel (d). The red boundaries in panels (c) and (d) highlight the difference in particulate penetration into the model, in the two cases. Note:~These images were created using FieldView\textsuperscript{TM}, as provided by Intelligent Light through its University Partners Program.}\label{f:inertial}
\end{center}
\vspace{-0.5cm}
\end{figure}

\begin{figure}[b]
\vspace{-0.15cm}
\begin{center}
\includegraphics[width=0.8\textwidth]{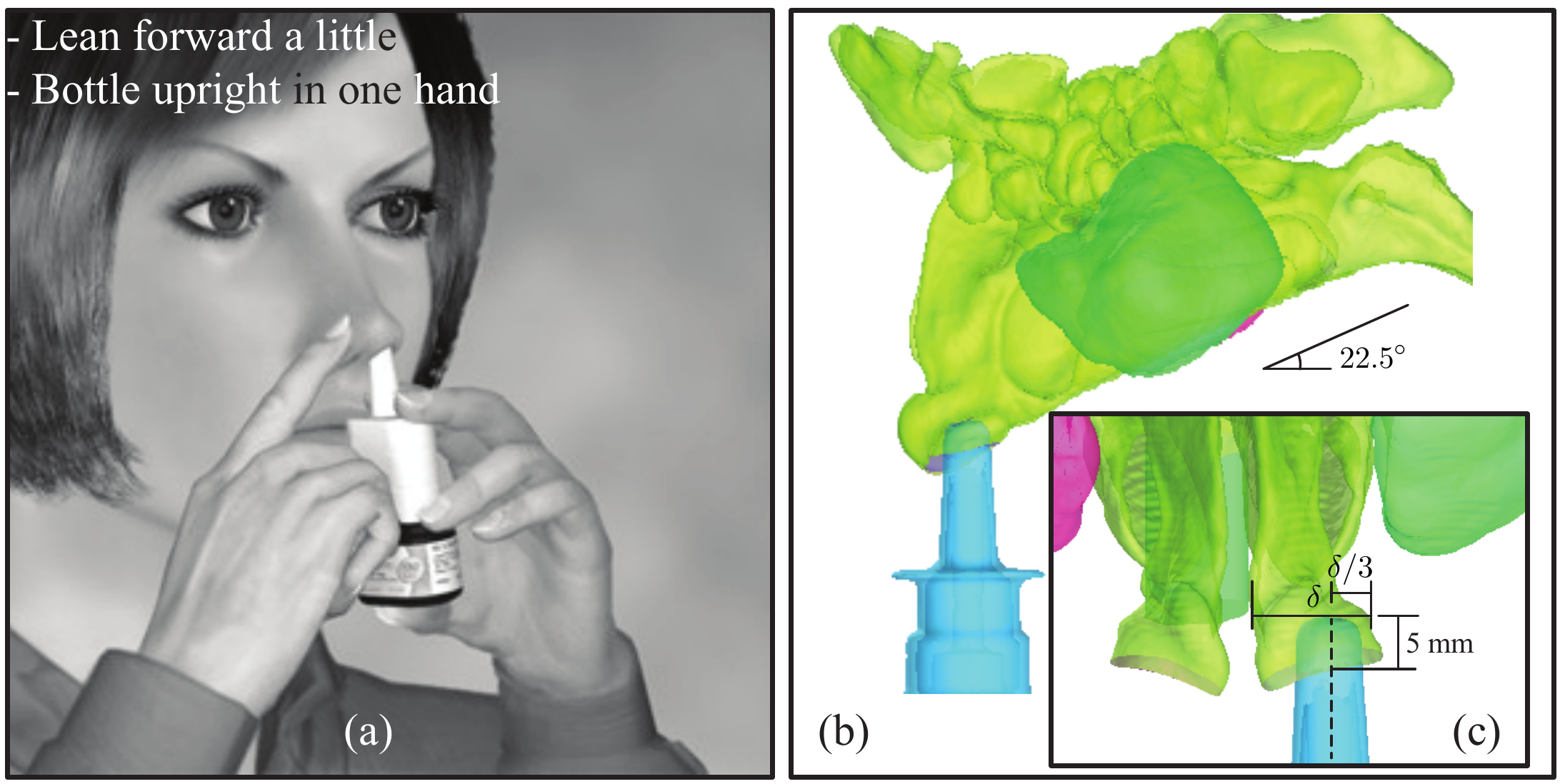}
\caption{(a) Sample pictorial usage instructions, available for over-the-counter nasal spray Flonase\textsuperscript{TM}; use of the graphic is subject to copyrights\cite{flonase2013}. Panel (b) and inset (c) depict the protocol implemented in the numerical simulations for the ``Current Use'' (CU) spray orientation. Note that $\delta$ is the linear distance between lateral wall and septum (the cartilaginous ``mid-wall'' in the nose, separating right and left airways) at 5-mm insertion depth into the nose. The model ``head'' is tilted forward by 22.5$^{\circ}$. The vertically upright dashed line represents the spray nozzle axis.}\label{f:cu}
\end{center}
\vspace{-0.5cm}
\end{figure}

\begin{figure}[t]
\vspace{-0.15cm}
\begin{center}
\includegraphics[width=1\textwidth]{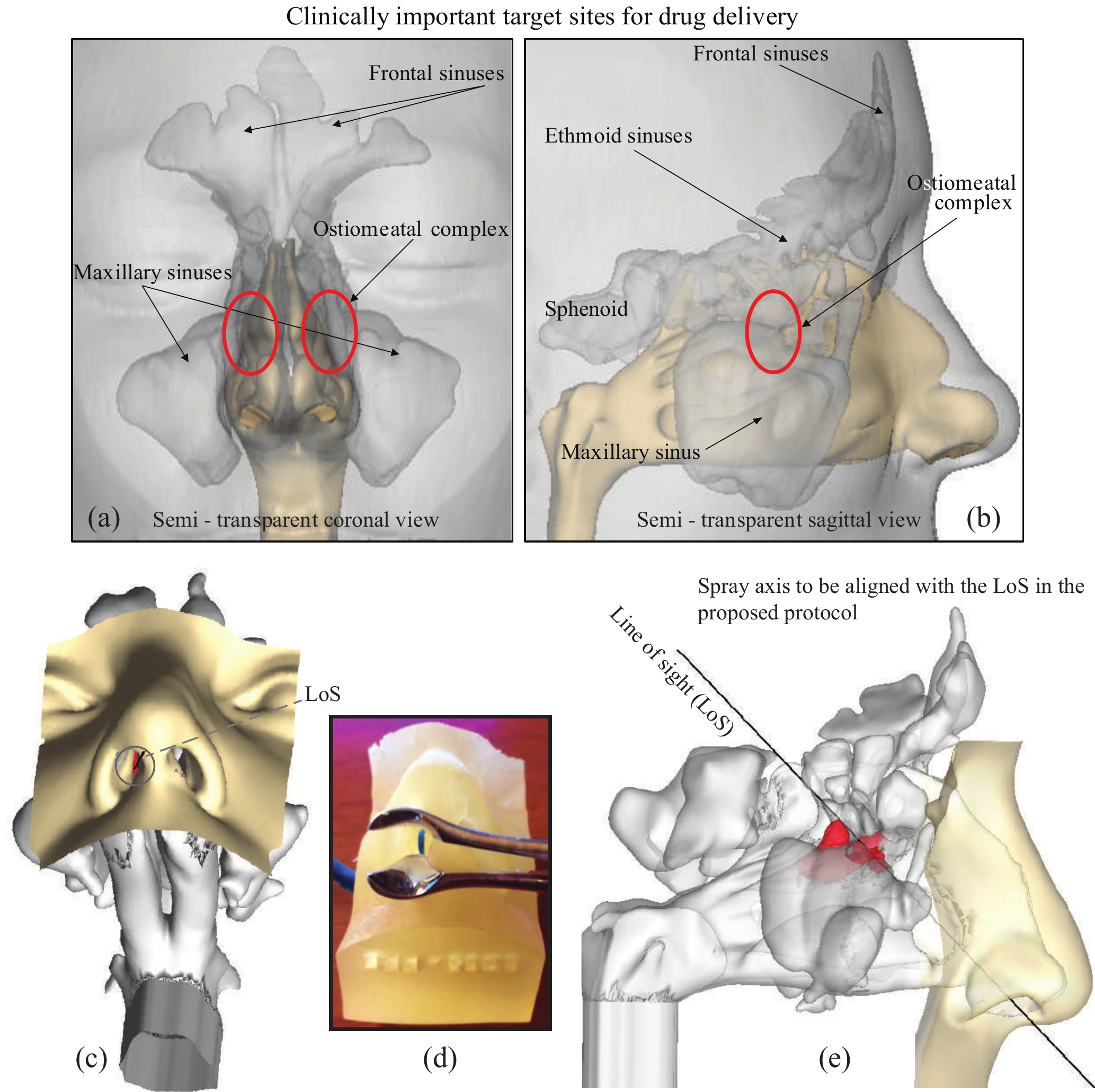}
\caption{Panels (a) and (b) show the locations of the main target sites in a representative sinonasal reconstruction, i.e.~the OMC (acting as the mucociliary drainage pathway for the sinuses) and the sinus cavities. Panels (c)-(e) demonstrate the ``Line of Sight'' (LoS; represented by the black lines) in NPM1. The anatomic zone, colored red, marks the OMC. Note that panel (d) is the 3D-printed soft nose from NPM1, exhibiting the same approximate orientation as that of the digital model in panel (c), giving a direct straight-line access to the target sites, and hence an LoS. The blue component in the image on panel (d) indicates the approximate location of the OMC.}\label{f:los}
\end{center}
\vspace{-0.5cm}
\end{figure}

\subsection{Identification of target site and spray parameters}\label{s:StudyDesign}

\subsubsection{Effect of airflow on droplet trajectories:}\label{s:FlowPhysics}
Inertial motion of a droplet is linearly proportional to its mass, and hence is exponentially proportional to the droplet diameter. Consequently, for bigger droplets, the inertial motion persists longer before being taken over by the ambient airflow. Figure~\ref{f:inertial}(a) tracks the trajectory of a representative 5$\mu$ droplet. In there, the tiny red circle marks the location where the inertial motion of the droplet got overwhelmed by the ambient flow, beyond which the droplet trajectory was same as the airflow streamline on which it was embedded at the red circle's location. Note the contrasting 25$\mu$ droplet trajectory in Figure~\ref{f:inertial}(b), where the inertial motion persisted longer. The phenomenon has a significant impact on drug deposition trends. The bigger droplets ($\ge$100$\mu$) show a greater propensity to hit the anterior walls directly owing to their high initial momentum, while smaller droplet sizes penetrate further into the airspace; see e.g.~Figure~\ref{f:inertial}(c)-(d). To ensure that the bigger droplets also reach the target sites, we argue that it would be conducive to harness their inertial motion and direct those droplets actively toward the target when they exit the spray nozzle. This can be feasibly achieved by orienting the spray axis to pass directly through an intended anatomic target zone.

\subsubsection{Current use instructions:}
Inconsistency and ambiguity in instructions\cite{benninger2004ohns, kundoor2011pr} indicate a lack of definitive knowledge on the best ways to use a nasal spray device. Different commercial sprayers often offer somewhat contrasting recommendations. However, there is a common agreement (see Figure~\ref{f:cu}(a)) that the patient should incline her/his head slightly forward, while keeping the spray bottle upright\cite{flonase2013, benninger2004ohns}. Furthermore, there is a clinical recommendation to avoid pointing the spray directly at the \textit{septum} (the separating cartilaginous wall between the two sides of the nose). These suggestions were adopted in our standardization\cite{kimbell2018rdd} of ``Current Use'' (CU) protocol for topical sprays. The digital models were inclined forward by an angle of 22.5$^{\circ}$, and the vertically upright\cite{benninger2004ohns} spray axis was closer to the lateral nasal wall, at one-third of the distance between the lateral side and septal wall. Also, the spray bottle was so placed that it penetrated into the airspace by a distance of 5-mm, inspired by the package recommendations of commercial sprayers\cite{flonase2013} for a ``shallow'' insertion into the nose. Figure~\ref{f:cu}(b) lays out the schematics of the CU protocol used in this study.

\subsubsection{Target site identification and proposing an alternate spray use criteria:}
All sinuses, except sphenoid, drain into the ostiomeatal complex (OMC), it being the main mucociliary drainage pathway and airflow exchange corridor between the nasal airway and the adjoining sinus cavities. To ensure that as many drug particulates reach the sinus chambers and their vicinity as would be possible, we hypothesize that the spray axis should be directed straight toward the OMC. This is supported by our observation of the effect of airflow physics on droplet trajectories (see discussion in Section~\ref{s:FlowPhysics}). If the spray axis hits the OMC directly, the likelihood that the larger droplets will deposit there is higher. We refer to this usage protocol as ``Line of Sight'' (LoS). Like the CU protocol, the LoS protocol also had the sprayer inserted at a depth of 5-mm into the nasal airspace. Representative LoS orientation is shown in Figure~\ref{f:los}.

TSPD percentage at the OMC and the sinuses was evaluated as $= 100 \times \left(M_{\textrm{target}}/M_\textrm{spray}\right)$;
with $M_{\textrm{target}}$ being the spray mass of the particulate droplets deposited at the OMC and inside the sinus cavities, and $M_\textrm{spray}$ being the mass of one spray shot. 

\subsubsection{Generation of varying peripheral directions around the true CU and LoS directions:}
To establish the robustness of the TSPD predictions for the CU and LoS protocols, we also tracked droplet transport and deposition when the spray directions were slightly perturbed. Such perturbed peripheral directions for CU initiated 1 mm away on the nostril plane and were parallel to the CU's vertically upright true direction. For LoS, the perturbed peripheral directions were obtained by connecting the base of the true LoS direction on the nostril plane with points that radially lie 1 mm away from a point on the LoS; this specific point being 10 mm away along the LoS from the base of the LoS direction on the nostril plane (e.g.~see bottom panel of Figure~\ref{fig:CUvsLOS} for an illustrative example).

\subsubsection{Parameters for the simulated spray shot:}
Over-the-counter Nasacort\textsuperscript{TM}~(Triamcinolone Acetonide), a commonly prescribed and commercially available nasal spray, was selected for this study. Four units of Nasacort\textsuperscript{TM}~were tested at Next Breath, LLC (Baltimore, MD, USA) to evaluate the \textit{in vitro} spray performance. Corresponding plume geometry was analysed through a SprayVIEW\textsuperscript{\textregistered} NOSP, which is a non-impaction laser sheet-based instrument. Averaged spray half-cone angle was estimated at 27.93$^\circ$, and the droplet sizes in a spray shot followed a log-normal distribution. With the droplet diameter as $x$, the droplet size distribution can be framed as a probability density function of the form\cite{cheng2001jam}:
\begin{equation}
f(x) = \frac{1}{\sqrt{2\pi}x \ln \sigma_{g}} \exp \left[ -\frac{(\ln x - \ln x_{50})^2}{2 (\ln \sigma_g)^2}  \right].
\end{equation}
Here, $x_{50} = 43.81\mu$ is the mass median diameter (alternatively, the geometric mean diameter \cite{finlay2001book}) and $\sigma_g = 1.994$ is the geometric standard deviation. The latter quantifies the span of the droplet size data. Measurements were also made with and without the saline additive in the sprayer, and the tests returned similar droplet size distribution. Note that a saline additive was used during the physical recording of the sprayed deposits. The mean spray exit velocity from the nozzle was 18.5 m/s, based on phase doppler anemometry-based measurements\cite{liu2011}. 

While simulating the droplet trajectories, we assumed typical solid-cone injections and tracked the transport for 1-mg spray shot while comparing the TSPD trends from the CFD predictions with the corresponding experimental drug delivery patterns.
On the other hand, 95.0306 mg (which is one shot of Nasacort\textsuperscript{TM}, as quantified by Next Breath, LLC) of spray mass transport was simulated while comparing the CFD-based TSPD numbers for the LoS and CU protocols in each model.


\begin{table}[t]
\caption{Numerical prediction of targeted drug delivery from CU and LoS protocols. The LoS TSPD values that are significantly higher than the corresponding CU TSPD are marked by `$^*$'. \textbf{Symbols:}~$\mathbf{\sigma} = $ standard deviation, $\mathbf{\mu} =$ mean.}\label{Table1}
\vspace{-0.35cm}
\begin{center}
\includegraphics[width=1.0\textwidth]{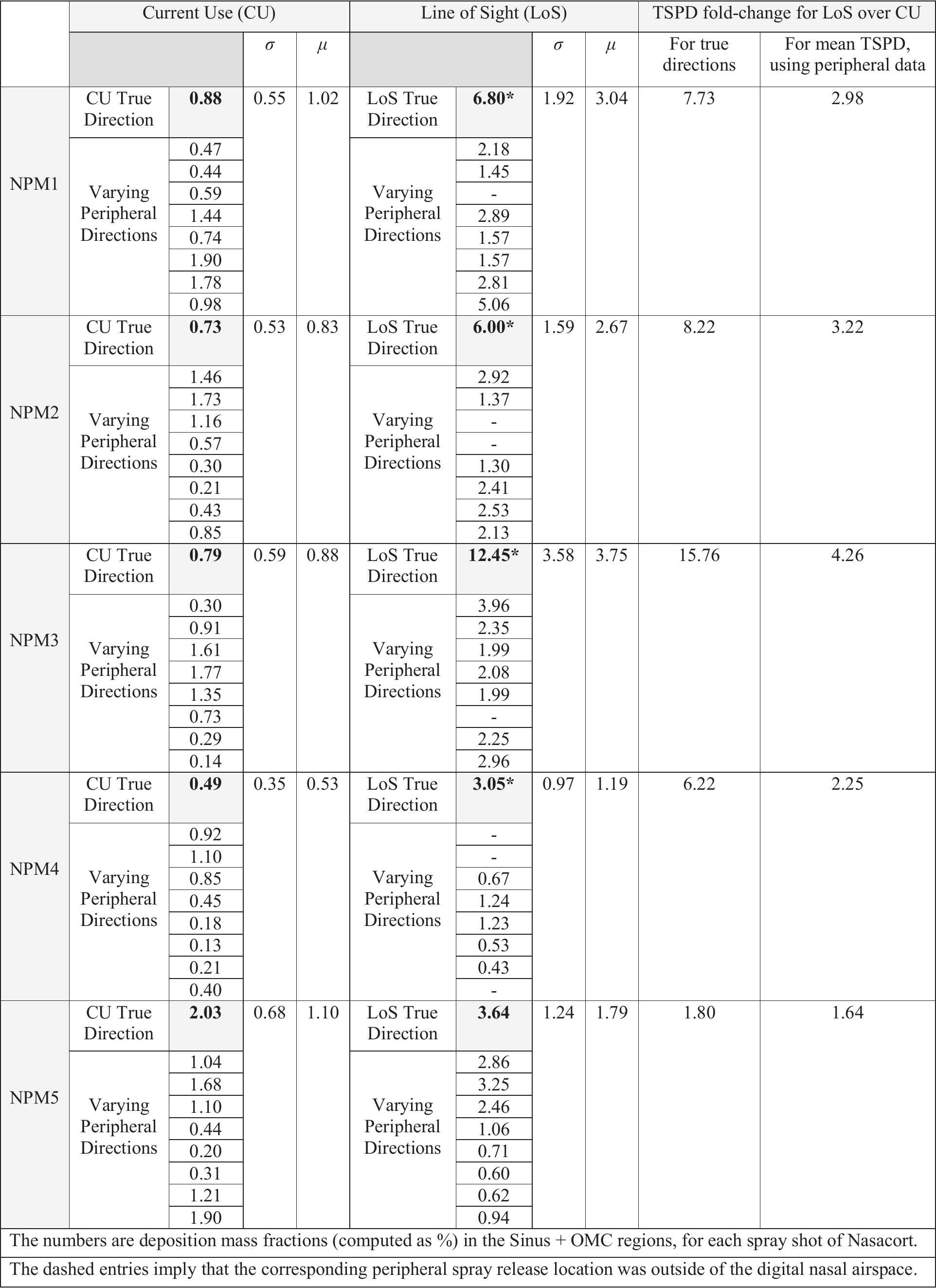}
\end{center}
\vspace{-0.5cm}
\end{table}


\section{Results}

\subsection{Comparison between CU and LoS spray usage protocols}

\begin{figure}[t]
\vspace{-0.15cm}
\begin{center}
\includegraphics[width=1.0\textwidth]{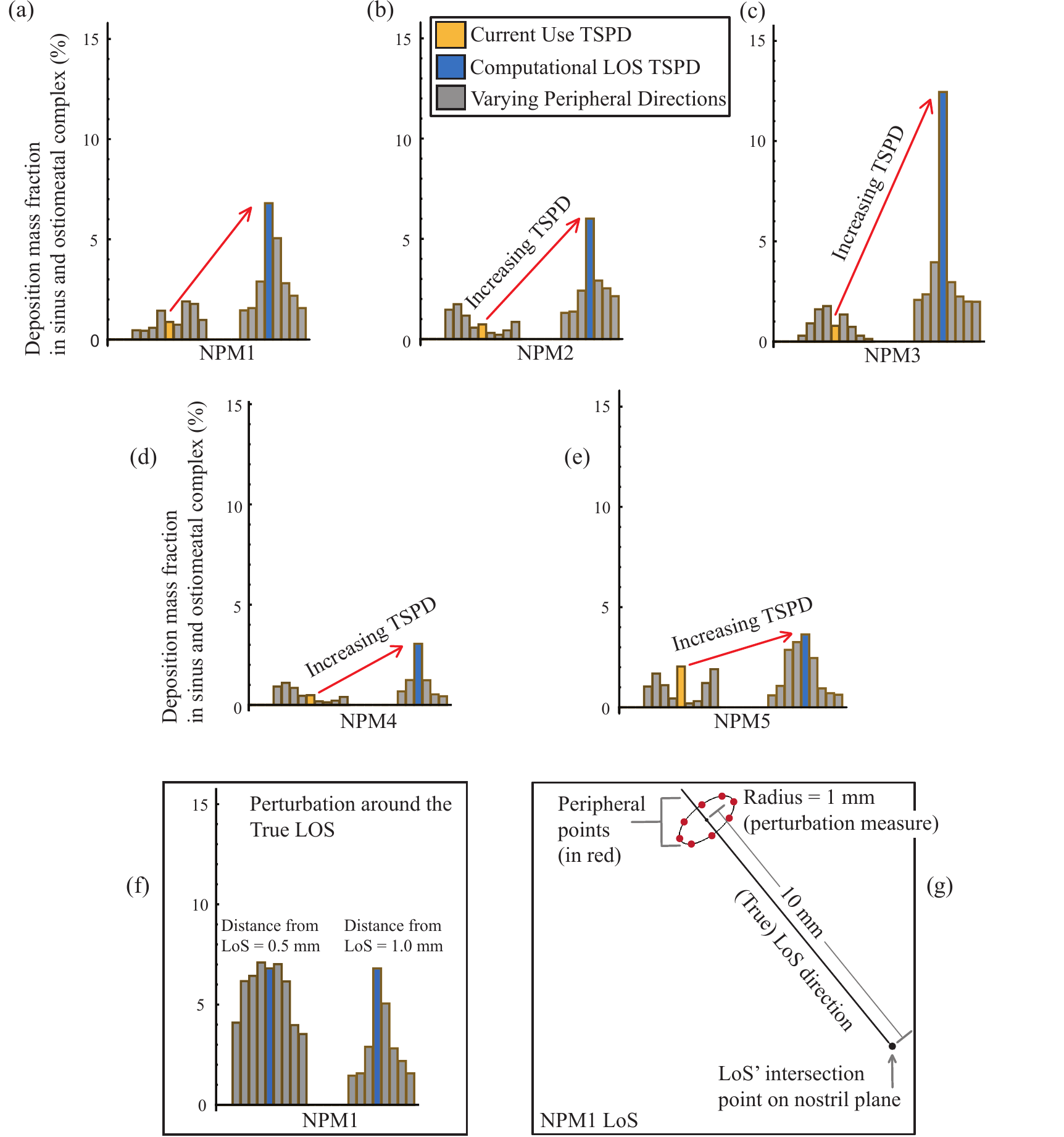}
\caption{Comparison of the simulated spray deposits from the CU and LoS protocols. The yellow bars represent the TSPD for the CU spray orientations, and the blue bars quantify the TSPD recorded for the LoS spray orientations. The gray bars are the predicted deposits when the true CU and LoS directions were perturbed by 1 mm. Panels (a)--(e) are the results for five different airway models: Nasal Passage Model 1 (NPM1), Nasal Passage Model 2 (NPM2), Nasal Passage Model 3 (NPM3), Nasal Passage Model 4 (NPM4), and Nasal Passage Model 5 (NPM5). Panel (f) compares the TSPD for peripheral directions in a 0.5-mm perturbation (on the left) with respect to a 1-mm perturbation (on the right) from the true LoS orientation, both in NPM1. As expected from the overall findings, the TSPD increased for the perturbed spray directions that were closer to the true LoS. Panel (g) depicts the spatial perturbation parameters for the LoS spray axis orientation in NPM1.}\label{fig:CUvsLOS}
\end{center}
\vspace{-0.5cm}
\end{figure}

LoS was found to be consistently superior in comparison to the CU spray placement protocol, while targeting the OMC and the sinus cavities for drug delivery. Table~\ref{Table1} lists the deposition fraction percentages for each spray release condition in the five airway models (NPM1 -- NPM5). For a graphical interpretation, we have plotted the same information on Figure~\ref{fig:CUvsLOS}. Overall, the deposition fraction for the LoS was on an average 8.0-fold higher than the CU deposition fraction, with the corresponding subject-specific improvement range being 1.8 -- 15.8 folds for the five test models. The improvement does decay when the perturbed peripheral spray directions are compared, to assess the robustness of the LoS protocol's advantage over CU. Considering the varying peripheral directions around the true LoS and CU, the LoS set registered an average 3.0-fold increase in TSPD, with the corresponding subject-specific improvement range being 1.6 -- 4.3 folds.

\subsubsection{Statistical tests -- on improvements achieved by the revised spray use strategy:}
LoS was compared to CU through a paired study design on the data from five test models. Table~\ref{Table2} lays out the computed numbers. For each model, the outcome comprised the percentage of deposition in OMC and the sinuses for both CU and LoS spray usage. Null hypothesis considered for this statistical test assumed that the TSPD would be same for CU and LoS in an airway model. The deposition percentage corresponding to CU and LoS protocols in the same nostril were treated as paired observations for a paired t-test to check the null hypothesis. Owing to a relatively small study cohort, paired Wilcoxon signed rank test was also used for robustness check. In order to study how spatial variation might affect the difference between CU and LoS, three different ways of calculating the percentage of deposition were implemented. The first strategy considered the average deposition from the true LoS and CU directions. The second strategy compared the TSPD averaged from the true CU and LoS directions, along with the deposition data for spray release parameters obtained by perturbing the respective true directions. The third strategy used TSPD averaged exclusively from the deposition data corresponding to the perturbed spray release parameters. This allowed us to assess the robustness of any probable improvement from using LoS, while still accounting for slight spatial variations of the spray direction.

\begin{table}[b]
\caption{Statistical tests for the comparison between CU and LoS protocols.}\label{Table2}
\vspace{-0.35cm}
\begin{center}
\includegraphics[width=1.0\textwidth]{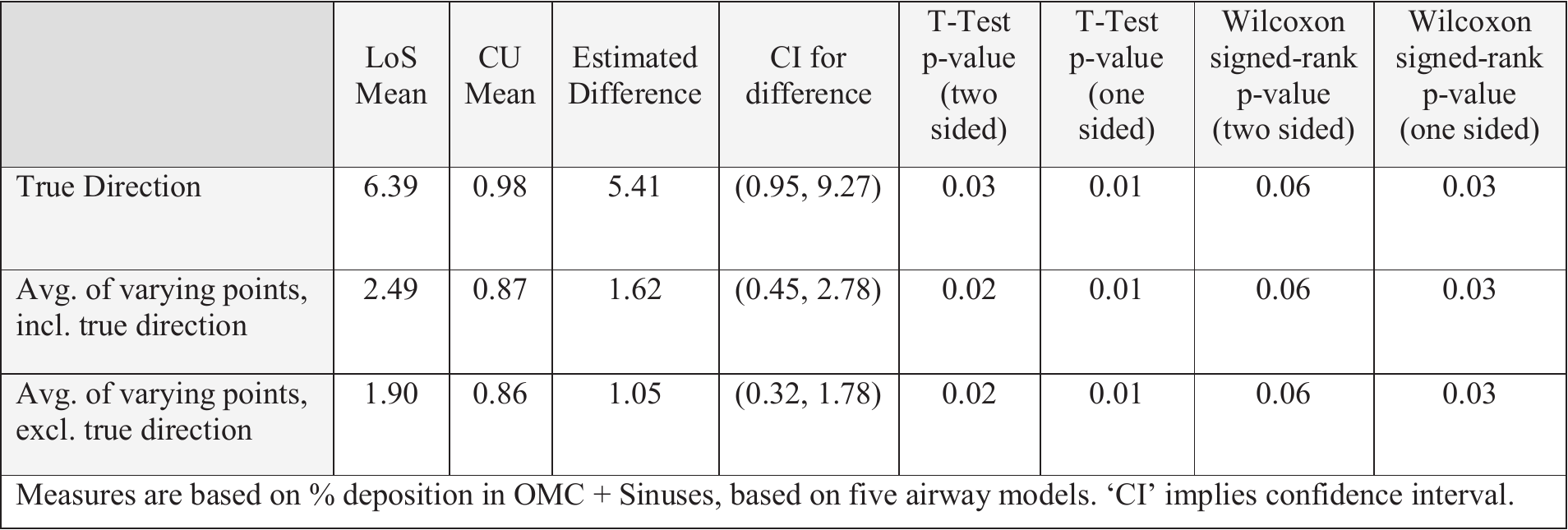}
\end{center}
\vspace{-0.5cm}
\end{table}

The first comparison method demonstrates an average deposition increase of 5.4 percentage points for LoS (6.39-\% for LoS vis-\`{a}-vis 0.98\% for CU). This difference is significant at the 0.05 level with a p-value from the paired t-test of 0.03. The paired Wilcoxon signed-rank test has a p-value of 0.06, which was the lowest possible p-value for the Wilcoxon signed-rank test given only five pairs of data. In the second comparison scheme, LoS has an increased deposition of 1.62 percentage points relative to CU (2.49\% vis-\`{a}-vis 0.87\%). The p-value for this difference is 0.02 using the paired t-test and 0.06 using the Wilcoxon signed rank test. Finally, for the third comparison method, LoS registered an increased deposition of 1.05 percentage points relative to CU (1.90\% vis-\`{a}-vis 0.86\%). The p-value for this difference is 0.02 using the paired t-test and 0.06 using the Wilcoxon signed rank test. This provides a strong evidence that LoS leads to higher percentage of deposition in the OMC and sinuses. The estimated difference is largest when using just the true directions, but the difference is still statistically significant even when using the spray release points obtained by perturbing the true directions. The p-value from the paired t-test is actually lower when the TSPD from just the perturbed points are considered, owing to the reduced variance for the estimated difference. For all three ways of estimating the percentage of deposition, the paired Wilcoxon signed-rank test returns a p-value of 0.06. With only five pairs of data, this suggests that the use of LoS does result in statistically significant higher deposition for all five nostril models.


\subsection{Comparison of the simulated TSPD predictions with physical experiments}
\begin{figure}[t]
\vspace{-0.15cm}
\begin{center}
\includegraphics[width=1.0\textwidth]{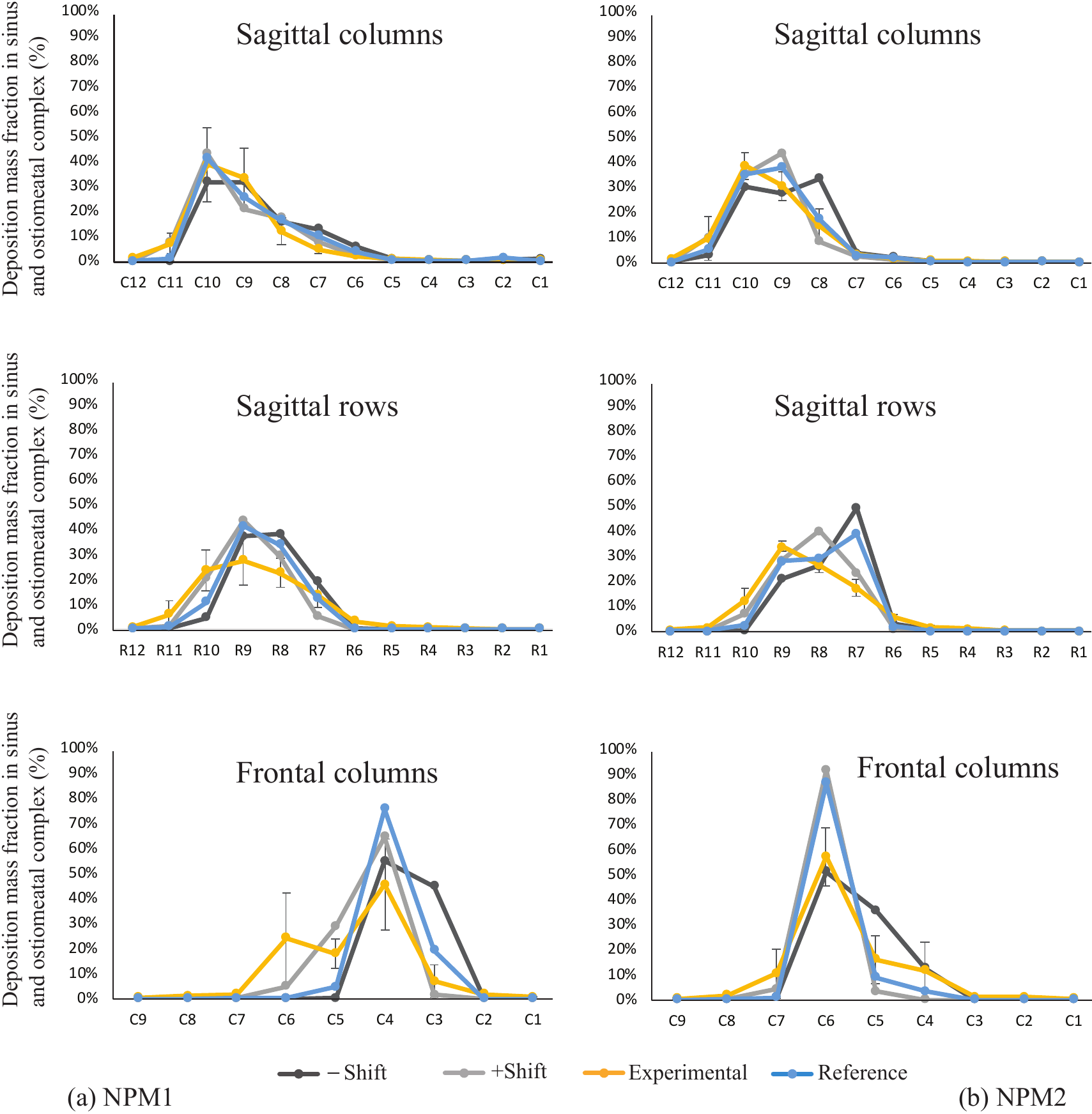}
\vspace{-0.25cm}
\caption{(a) Comparison of the numerically simulated compartmental findings in Nasal Passage Model 1, with respect to the gamma scintigraphy recordings from the corresponding 3D-printed replica. (b) Comparison of the numerically simulated compartmental findings in Nasal Passage Model 2, with respect to the gamma scintigraphy recordings from the corresponding 3D-printed replica. The blue ``reference'' lines trace the CFD predictions for TSPD in each compartment, with the light gray and dark gray lines respectively marking the variability in prediction, for +/- 1 pixel shift while superimposing the gridlines on the numerical data-space. The yellow lines trace the TSPD recorded from the physical experiments.
}\label{fig:Comparison}
\end{center}
\vspace{-0.6cm}
\end{figure}

Figure~\ref{fig:Comparison} compares the numerical TSPD predictions with corresponding gamma scintigraphy-based experimental recordings in NPM1 and NPM2. While the compartmental deposits visibly presented a congruous trend in the sagittal columns, sagittal rows, and frontal columns; we conducted additional statistical tests to verify the homogeneity between the two sets of data so as to establish the reliability of the computational findings.

\begin{table}[b]
\caption{Comparison between the compartmental data from numerical simulations and physical experiments.}\label{Table3}
\vspace{-0.35cm}
\begin{center}
\includegraphics[width=1.0\textwidth]{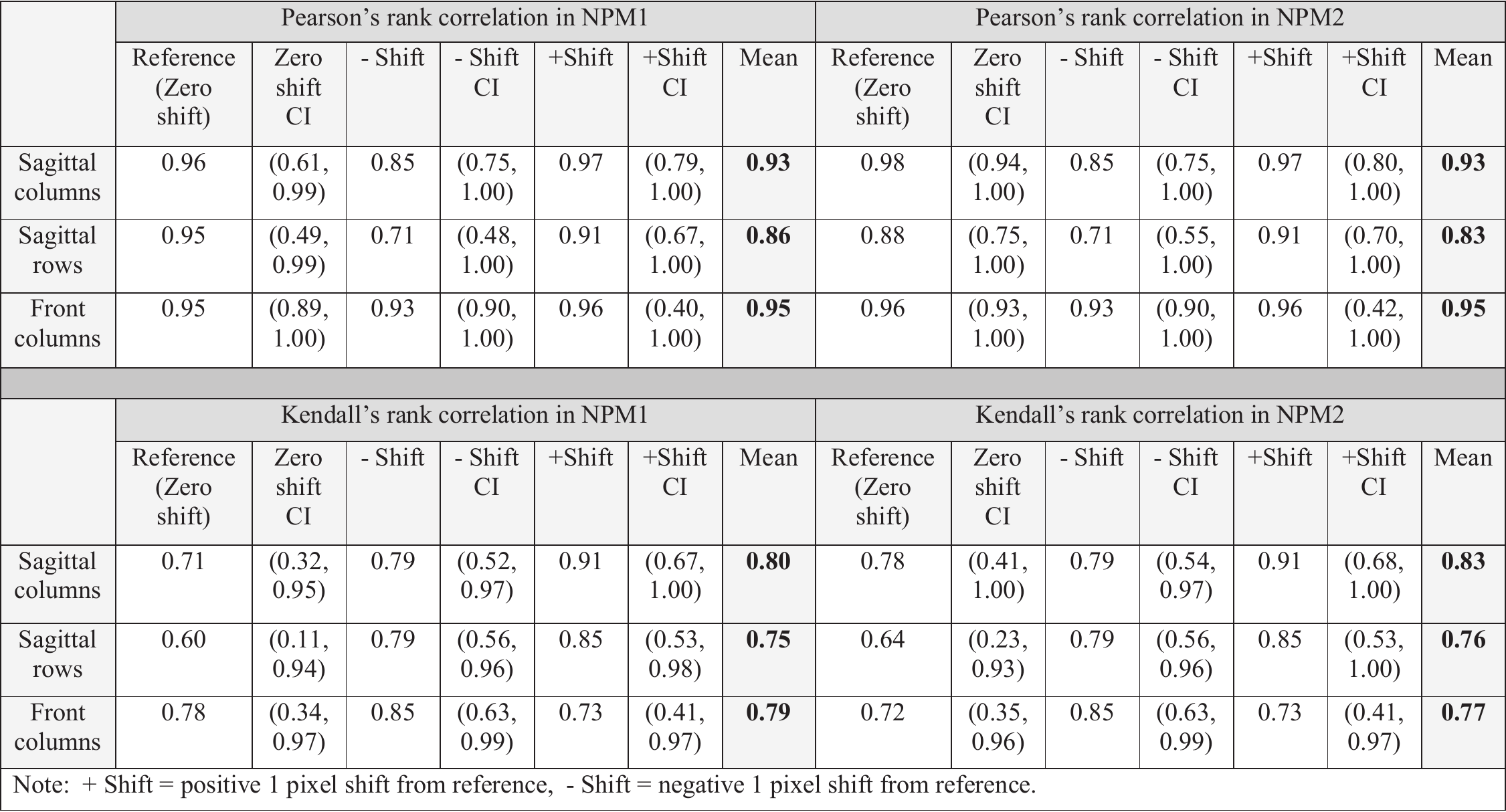}
\end{center}
\vspace{-0.5cm}
\end{table}

Table~\ref{Table3} gives the Pearson and Kendall’s correlation between the numerical and experimental models for the average deposition fractions in NPM1 and NPM2 for the LoS protocol. The confidence intervals are based on 1000 bootstrap samples, instead of asymptotic approximations, because of the relatively small sample size. Based on the output, we can see that the Pearson correlation is consistently very high while the Kendall’s correlation is somewhat lower. However, while the Kendall’s correlation is frequently thought to be more robust to outliers, particularly for small sample sizes like this data-set; in this particular instance the Pearson correlation is likely more illustrative. This is because the Pearson correlation is able to show that, for the most part, the magnitudes of the estimates are similar and comparable between the numerical and experimental models. In general, there is a strong linear relationship between the percent of deposition prediction from the numerical model and the corresponding physical measurements in the experimental model. The lower Kendall’s correlation (overall mean measure 0.78) is largely due to regions where both the numerical and experimental models had very low average deposition but the exact rank of these regions changed considerably between the two data-sets. Note that this does not necessarily indicate a poor performing numerical model. However, the relatively high Pearson correlation (overall mean measure 0.91) does indicate that the numerical models perform well while predicting the sprayed droplet transport.\\




\section{Discussion}
CFD-guided nasal spray usage defined by the LoS protocol was found to significantly enhance topical drug delivery at targeted sinonasal sites, when compared to currently used spray administration techniques. With increased sample size, this work can be the catalysis toward prompting personalized instructions and specifications for improved use of topical sprays. The findings, thus, have the potential to substantially upgrade the treatment paradigm for sinonasal ailments through the ability to ascertain LoS in individual subjects via endoscopic examinations conducted in the clinic, and to help guide treatment decision-making and patient instructions for spray usage.

\begin{table}[b]
\caption{Comparison of the LoS scores, obtained observationally and through determining the surface area projection of the targeted OMC on the nostril plane.}\label{Table4}
\vspace{-0.35cm}
\begin{center}
\includegraphics[width=1.0\textwidth]{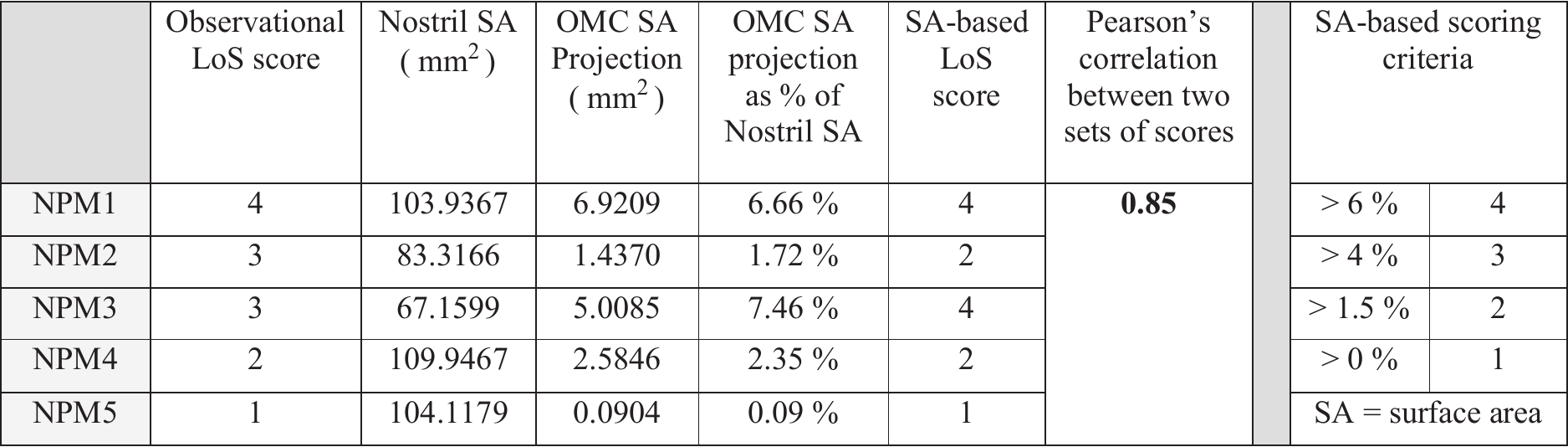}
\end{center}
\vspace{-0.5cm}
\end{table}

\subsection{Concept of LoS scoring and on the adaptability of our findings in clinical practice}
As means to quantifying the suitability of a person's airway for the LoS spray protocol, we exploratorily propose a scoring system that is based on how much of the targeted drug delivery sites (OMC, sinuses) are visible when inspected clinically from outside of the nostril. The scoring system will also serve to quantify nasal anatomic variability among individuals. Accordingly, as part of the current study, the LoS scores (see Table~\ref{Table4}) were first determined observationally, based on the external visibility of the OMC site in the \textit{in silico} sinonasal reconstructions. We fixed a range of scores $\in [\,1$, 4$\,]$, with 4 being used when the LoS direction was easiest to ascertain. Subjective as that scoring procedure may be, it is similar to what attending physicians will gauge during a clinic visit to determine if a particular patient has a ``line of sight'' in her/his nasal anatomy. So, to establish the relevance of the findings from this manuscript toward revisions of the therapeutic protocol for sinonasal care, it is important to assess the comparability of the observational LoS scores with more objective score determination techniques. 
This was achieved by calculating the surface area of the nostril plane and the projected area of the OMC on the plane of the nostril. We computed the ratio of the projected area to the nostril area, as a percentage. Scores of 4 were assigned if the ratio exceeded 6\%, 3 if the ratio exceeded 4\%, 2 if the ratio was more than 1.5\%, and 1 if the ratio was greater than 0\%. The two scoring techniques yielded very similar results (see Table~\ref{Table4}), with the highest and lowest scores respectively going to the same anatomic models. Pearson's rank correlation for the two sets of scores was 0.85. While a broader study, involving clinical trials, will be necessary to revise therapeutic protocol for nasal drug delivery, the present results illustrate the easy adaptability of our findings into clinical practice settings.

\subsection{On the comparability of the experimental data with the numerical findings}
The computational simulations assumed a laminar framework to mimic steady breathing. However, one may argue that even with resting breathing rates, the airflow often contains transitional features like vortices, emerging from the roll-up of shearing fluid layers during flow-structure interactions\cite{stremler2014fdr, basu2017jfm2} at the anatomic bends. Some of these nuances are, in fact, difficult to model without proper turbulence simulations\cite{zhao2014ifar, calmet2019plos}. However, true as that may be, the effect of these flow artifacts on eventual drug delivery in the sinuses has been found to be somewhat nominal while comparing laminar and turbulence simulation results\cite{basu2017num}. 

On the other hand, the \textit{in vitro} techniques also often pose challenges. For instance, there can be post-deposition run-off as the deposited solution traces undergo translocation along the inner walls of the solid replica. Such drip-off dynamics can lead to a flawed estimate of regional deposition. 

In the gamma scintigraphy-based method of recording deposits, the radiation signal undergoes some level of scattering and hence in the process of signal extraction from each of the compartments, there is the possibility that signals from one compartment may contaminate the signals at neighboring compartments. To minimize this effect while carrying out the comparisons, the nose (the soft plastic anterior part in the 3D-printed models), which had a bright radiation signal owing to the relatively large amount of anterior deposits, was excluded from both the experimental and numerical data.

Finally, while the inhalation airflow rates were same \textit{in vitro} and \textit{in silico}, the airflow partitioning on the two sides of the nasal airways was likely affected by the placement of the NPD, while administering the spray through hand-actuation.

\subsection{Caveats and future implications}
Readers should note that this was a computational study with validation from spray transport observations in inanimate solid replicas. Also, not every patient will have a clear access to the OMC, and hence may be \textit{without} an LoS. For instance, in the current study, of the six airway sides in the three study subjects, subject 2's right-side airway did not exhibit an LoS. 

This study, its restricted sample size and limitations notwithstanding, is, to the best of our knowledge, the first-of-its-kind to propose an alternative \textit{easy-to-implement} strategy that can significantly improve the intra-nasal delivery of topical drugs at the diseased sites. The recommendation for using the ``line of sight'' is user-friendly, personalized (the physician can instruct the patient on the spray usage technique based on a fast LoS check in the clinic), and has the potential to be smoothly incorporated into the nasal standard-of-care. For probable revisions to the clinical regimen, we will need a broader study with more subjects, along with a component for clinical trials to track patient response. Comparison of the numerical data with \textit{in vivo} spray performance will also eliminate errors that contaminate the \textit{in vitro} TSPD numbers (e.g.~from drip-off of the deposited solution along the inner wall contours of the 3D-printed models). Nevertheless on a larger intriguing perspective, the current study conclusively postulates how relatively simple engineering analysis and mechanistic tools can usher in transformative changes in the prognosis and treatment protocol for common ailments like nasal congestion. 

\noindent\hrulefill


\vspace{-0.25cm}

\section*{Acknowledgements}
The authors sincerely thank Dr.~John S Rhee, MD, MPH (at the Department of Otolaryngology, Medical College of Wisconsin) for numerous fruitful discussions. Thanks are also due to Dr.~Julie Suman (Next Breath, LLC) for the experimental measurement of nasal spray parameters. The authors additionally acknowledge: (a) Christopher Jadelis (at UNC Chapel Hill) for his assistance on the experimental setup; (b) several past/present UNC rhinology residents  and fellows (Drs.~Andrew Coniglio, Satyan Sreenath, Kibwei McKinney, Gita Madan, Parth Shah, and Stan McClurg) for their inputs; and (c) Dr.~Ola Harrysson’s group at NC State University (at the Edward P Fitts Department of Industrial and Systems Engineering), Matthew White (at NCSU), and Dr.~Tim Horn (Director of Research, Center for Additive Manufacturing and Logistics at NCSU) for help on 3D printing. Finally, thanks are also due to Alison Turner and Carolyn Hamby (both at UNC School of Medicine) for their assistance in patient recruitment scheduling.

Preliminary results pertaining to this work have featured at the American Physical Society (APS) -- Division of Fluid Dynamics Annual Meetings \cite{basu2018aps, basu2017aps} and at the International Society for Aerosols in Medicine (ISAM) Congress \cite{basu2019isam, farzal2019isam}.

The project was supported by:~(a)~the National Heart, Lung, and Blood Institute (NHLBI) of the National Institutes of Health (NIH), under award number R01HL122154 (PI:~JSK); (b)~the National Center for Advancing Translational Sciences at NIH, through award number KL2TR002490 (PI:~AJK); and (c)~SB's faculty start-up funds at the Department of Mechanical Engineering at South Dakota State University. Content of this study is solely the responsibility of the authors and does not necessarily represent the official views of the NIH.\\

\footnotesize
\noindent \textbf{Contributions:}~SB, GJMG, DOFI, BAS, AMZ, CSE, WDB, JPF, and JSK conceived this study; JSK led the patient recruitment with JW, ZF, MM, SB lending assistance; JSK, SB, ZF, MM developed the digital reconstructions; SB, JSK, OF, ZF ran the numerical simulations; LTH, JW, AB, WDB carried out the physical experiments; SB, KK, GJMG, DOFI, JSK post-processed the numerical and experimental data; JPF, BL, SB ran the statistical tests; BAS, AMZ, CSE, AJK, BDT, ZF, MM facilitated patient recruitment and provided clinical inputs; SB drafted the manuscript. Note that BAS, AMZ, CSE, AJK, BDT are attending physicians at the Division of Rhinology at UNC School of Medicine.\\
\normalsize

\noindent\textbf{Note: This is the pre-peer review version of the manuscript.}\\

\noindent\hrulefill

\noindent\textbf{References}\\

\bibliographystyle{unsrt}
\bibliography{NSFv10}

\end{document}